\documentclass[twoside,twocolumn]{article}

\usepackage{blindtext} 

\usepackage[sc]{mathpazo} 
\usepackage[T1]{fontenc} 
\usepackage{microtype} 

\usepackage[english]{babel} 

\usepackage[hmarginratio=1:1,top=32mm,columnsep=20pt]{geometry} 
\usepackage[hang, small,labelfont=bf,up,textfont=it,up]{caption} 
\usepackage{booktabs} 

\usepackage{lettrine} 

\usepackage{amsmath}
\usepackage{amssymb}
\usepackage{relsize}
\usepackage{graphicx}
\usepackage{amsthm}

\renewcommand{\vec}[1]{\boldsymbol{#1}} 

\usepackage{enumitem} 
\setlist[itemize]{noitemsep} 

\usepackage{abstract} 

\usepackage{titlesec} 
\renewcommand\thesection{\Roman{section}} 
\renewcommand\thesubsection{\roman{subsection}} 
\titleformat{\section}[block]{\large\scshape\centering}{\thesection.}{1em}{} 
\titleformat{\subsection}[block]{\large}{\thesubsection.}{1em}{} 

\usepackage{fancyhdr} 
\usepackage{titling} 

\usepackage{hyperref} 

\pretitle{\begin{center}\Huge\bfseries} 
	\posttitle{\end{center}} 
\title{Trajectories of two identical particles on a plane in a constant magnetic field subject to non-Coulomb potentials} 
\author{%
	\textsc{Malik Amir} \\[1ex] 
	\normalsize McGill University \\ 
	\normalsize \href{mailto:malik.amir@mail.mcgill.ca}{malik.amir@mail.mcgill.ca} 
	\and 
	\textsc{Andr\'e Valli\`eres} \\[1ex] 
	\normalsize McGill University \\ 
	\normalsize \href{mailto:andre.vallieres@mail.mcgill.ca}{andre.vallieres@mail.mcgill.ca} 
}
\date{\today} 
\renewcommand{\maketitlehookd}{%
	\begin{abstract}
		\noindent  
		Study of the classical motion of two identical particles on a plane subject to non-Coulomb potentials in a constant magnetic field presented in polar coordinates. With the rigorous analysis of the potentials and the constants of motion, we are able to give necessary and sufficient conditions for the existence of bounded and periodic orbits. We also show and analyze numerical solutions of Newton equations of motion.
	\end{abstract}
}

\begin{document}
	
	\maketitle
	
	
	\section{Introduction}
	
	\lettrine[nindent=0em,lines=3]{I} n quantum mechanics, it is often pointless to study trajectories because the focus of attention is on states and energy rather than position or time. However, sometimes we would like to have an idea of what a particle's trajectory looks like or at least, where it is heading to for a given potential. In this case, it is useful to approach the problem with classical mechanics, assuming that the particles behave classically and that their velocities and accelerations are small enough to neglect any relativistic or radiation effects.
	
	In this paper, we study the planar motion of two identical charges moving in a constant and uniform magnetic field subject to a non-Coulomb interaction. Our goal is to provide complete trajectory's information in order to guide further understanding of potentials which permit exact solutions in quantum mechanics. The three potentials we explore are treated in \cite{Kreshchuk:2015} in order to extend the class of quantum mechanics problems which permit quasi-exact solutions.
	
	We also continue the work of \cite{CurilefClaro:1995} by providing a complete method to the study of trajectories based on the derivation of the equations of motion and the constraints given by the constants of motion. These are, as seen later, the Hamiltonian and the angular momentum. In addition, we present a complete analysis of Kreshchuk's potentials in order to establish a relation between the constants composing them, so that we can set up restrictions on their possible values since these are not to be chosen arbitrarily to obtain a bounded or a periodic motion. This analysis will take place based on the behaviour of the effective potential.
	
	Since we focus our study on identical charges, the coupling charge is null and the problem is then separable in a center of mass motion and a relative motion. The center of mass moves in a constant magnetic field as a free particle of twice the charge and mass of each constituent of the pair of particles. The relative motion is in turn, that of a particle of half the charge and mass of each constituent. It moves in the presence of a constant magnetic field and an electric field produced by a particle fixed at the origin of twice the charge of each constituent of the pair. In further sections, all three potentials will be treated the same way using polar coordinates. We will present each step of our analysis of the first potential in section \ref{potentials}.\ref{1st-potential} and will refer the reader to this section while exploring the two other potentials in sections \ref{potentials}.\ref{2nd-potential} and \ref{potentials}.\ref{3rd-potential} where similar calculations occur. 
	
	
	\section{General solution}
	
	The Lagrangian which describes two non-relativistic particles $(q_1, m_1), (q_2, m_2)$ on a plane subject to a constant and uniform magnetic field $\vec{B}=B\vec{\hat{z}}$ perpendicular
	to the plane is of the form
	\begin{equation}
	\label{eq:lagr-gen-u}
	\begin{split}
	\mathcal{L}&=\frac{1}{2}m_1\vec{\dot{\rho}_1}^2+\frac{1}{2}m_2\vec{\dot{\rho}_2}^2\\
	&+\frac{q_1}{c}\vec{A}(\vec{\rho_1})\cdot\vec{\dot{\rho_1}}+\frac{q_2}{c}\vec{A}(\vec{\rho}_2)\cdot\vec{\dot{\rho_2}}\\
	&-V
	\end{split}
	\end{equation}
	where $\vec{A}$ is the vector potential, $m_i$, $q_i$, $\vec{\rho_i}$, $V$ are respectively the mass, the charge, the relative position vector and the general potential depending only on the relative distance between the two particles. Since we focus on identical particles, we can define $m \equiv m_1 = m_2$ and
	$q \equiv q_1 = q_2$. Using Gaussian units, the Lagrangian can now be rewritten as
	\begin{equation}
	\label{eq:lagr-u}
	\begin{split}
	\mathcal{L}&=\frac{1}{2}m\vec{\dot{\rho}_1}^2+\frac{1}{2}m\vec{\dot{\rho}_2}^2\\
	&+q\vec{A}(\vec{\rho_1})\cdot\vec{\dot{\rho_1}}+q\vec{A}(\vec{\rho_2})\cdot\vec{\dot{\rho_2}}\\
	&-V
	\end{split}
	\end{equation}
	Defining $\vec{R}=\frac{1}{2}(\vec{\rho_1}+\vec{\rho_2})$ as the position of the center of mass and $\vec{\rho}=\vec{\rho_2}-\vec{\rho_1}$ as the
	relative position vector between the two particles, we obtain for the central mass motion
	\begin{equation}
	\label{eq:lagr-cm-u}
	\mathcal{L}_{cm}=m(\dot{R}^2+R^2\dot{\theta}^2_{cm})+qBR^2\dot{\theta}_{cm}
	\end{equation}
	while for the relative motion
	\begin{equation}
	\label{eq:lagr-rel-u}
	\mathcal{L}_{rel}=\frac{1}{4}m(\dot{\rho}^2+\rho^2\dot{\theta}_{rel}^2)+\frac{1}{4}qB\rho^2\dot{\theta}_{rel}-V
	\end{equation}
	An important point to make here is that $\mathcal{L}_{cm}$ is independent of the potential $V$; hence, the motion of the center of mass is the same for all three potentials. From these Lagrangians, we can derive the Hamiltonians with the conventional Legendre transformation
	\begin{equation}
	\label{eq:ham-cm-u}
	\begin{split}
	\mathcal{H}_{cm}&=\frac{1}{4}\frac{{p^{cm}_{\theta}}^2}{mR^2}+\frac{1}{4}\frac{p_R^2}{m}\\
	&-\frac{1}{2}\frac{qBp^{cm}_{\theta}}{m}+\frac{1}{4}\frac{q^2B^2R^2}{m}\\
	&=m(\dot{R}^2+R^2\dot{\theta}^2_{cm})
	\end{split}
	\end{equation}
	and
	\begin{equation}
	\label{eq:ham-rel-u}
	\begin{split}
	\mathcal{H}_{rel}&=\frac{{p^{rel}_{\theta}}^2}{m\rho^2}+\frac{p_{\rho}^2}{m}\\
	&-\frac{1}{2}\frac{qBp^{rel}_{\theta}}{m}+\frac{1}{16}\frac{q^2B^2\rho^2}{m}+V\\
	&=\frac{1}{4}m(\dot{\rho}^2+\rho^2\dot{\theta}_{rel}^2)+V
	\end{split}
	\end{equation}
	where $p_R=2m\dot{R}$ and $p_\rho=\frac{1}{4}(2m\dot{\rho})$ are respectively the linear momentum for the central motion and the relative motion, and
	$p^{cm}_{\theta}=2mR^2\dot{\theta}_{cm}+qBR^2$ and $p^{rel}_{\theta}=\frac{1}{4}(2m\rho^2\dot{\theta}_{rel}+qB\rho^2)$ are respectively the angular momentum for the central motion and the relative motion.
	
	
	\section{Central motion}
	
	The central motion trajectory is derived from (\ref{eq:lagr-cm-u}) to obtain the following Newton equations
	\begin{equation}
	\label{eq:newton-cm-u-1}
	m\ddot{R}=mR\dot{\theta}_{cm}^2+qBR\dot{\theta}_{cm}
	\end{equation}
	and
	\begin{equation}
	\label{eq:newton-cm-u-2}
	4mR\dot{R}\dot{\theta}_{cm}+2mR^2\ddot{\theta}_{cm}+2qBR\dot{R}=0
	\end{equation}
	To simplify the analysis, we introduce dimensionless parameters. Let us define $\xi=\frac{R}{\ell_B}$, $\dot{\xi}=\frac{\dot{R}}{\ell_B\omega_c}$, and $\ddot{\xi}=\frac{\ddot{R}}{\ell_B\omega_c^2}$, where $\ell_B=\sqrt[3]{\frac{m}{B^2}}$ and $\omega_c=\frac{qB}{m}$. We can easily verify that time, energy and angular momentum are expressed in units of $\frac{1}{\omega_c}$, $\frac{q^2}{\ell_B}$ and $m\omega_c\ell_B^2$ respectively. 
	From there, we can write the center of mass Lagrangian and Hamiltonian as follow
	\begin{equation}
	\label{eq:lagr-cm-nu}
	\mathcal{L}_{cm}=\xi^2(\frac{\dot{\theta}_{cm}^2}{\omega_c^2}+\frac{\dot{\theta}_{cm}}{\omega_c})+\dot{\xi}^2
	\end{equation}
	and
	\begin{equation}
	\label{eq:ham-cm-nu}
	\mathcal{H}_{cm}=\frac{1}{4}(\frac{p^{cm}_{\theta}}{\xi}-\xi)^2+\dot{\xi}^2
	\end{equation}
	where 
	\begin{equation}
	\label{eq:p_theta-cm}
	p_{\theta}^{cm}=2\xi^2(\frac{\dot{\theta}_{cm}}{\omega_c}+\frac{1}{2})
	\end{equation}
	is a constant of motion. This can be seen by noticing that $\theta$ does not appear in $\mathcal{L}_{cm}$ or that $\{p_\theta^{cm},\mathcal{H}_{cm}\}=0$.
	\newline
	
	It follows that, from $\mathcal{L}_{cm}$, the dimensionless Newton equations of central motion are
	\begin{equation}
	\label{eq:newton-cm-nu-1}
	\ddot{\xi}=\xi(\frac{\dot{\theta}_{cm}^2}{\omega_c^2}+\frac{\dot{\theta}_{cm}}{\omega_c})
	\end{equation}
	and
	\begin{equation}
	\label{eq:newton-cm-nu-2}
	2\xi\dot{\xi}\frac{\dot{\theta}_{cm}}{\omega_c}+\xi^2\frac{\ddot{\theta}_{cm}}{\omega_c^2}+\xi\dot{\xi}=0
	\end{equation}
	Based on \cite{CurilefClaro:1995}, the above yields the following equation of motion
	\begin{equation}
	\label{eq:motion-cm}
	\xi^2-2\xi\sqrt{\mathcal{H}_{cm}+p^{cm}_{\theta}}\cos(\theta_{cm}-\theta_0)+p^{cm}_{\theta}=0
	\end{equation}
	which describes a circle of radius $\sqrt{\mathcal{H}_{cm}}$ centered at $(\sqrt{\mathcal{H}_{cm}+p^{cm}_{\theta}}$, $0)$ for $\theta_0=0$. It is important to note that the motion describes a circle with angular frequency $\dot{\theta} = -\omega_c=-\frac{qB}{m}$, where $\omega_c$ is the well-known cyclotron frequency.
	
	
	\section{Motion for the Three Potentials} \label{potentials}
	
	\subsection{First potential} \label{1st-potential}
	
	The first potential has the form
	\begin{equation}
	\label{eq:v1-u}
	\boxed{V_1=\frac{a}{\rho}+\frac{b}{\rho^2}+c\rho+d\rho^2}
	\end{equation}
	
	\begin{figure}[h]
		\centering
		\resizebox{0.48\textwidth}{!}{
			\includegraphics{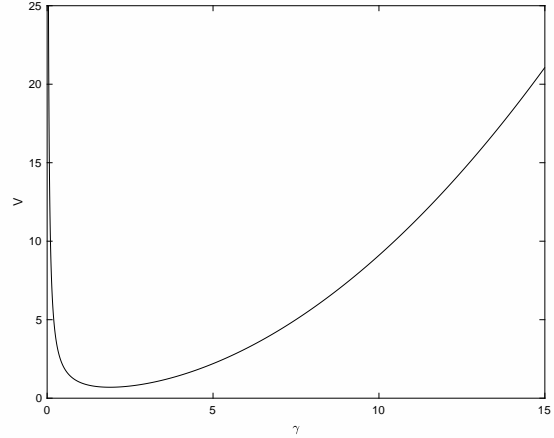}
		}
		\caption{Potential 1}
	\end{figure} 
	where $a=q_1q_2=q^2$, and $b,c,d \in \mathbb{R}$. As in the central motion case, we introduce dimensionless parameters to simplify the analysis. We thus define $\gamma=\frac{\rho}{\ell_B}$, $\dot{\gamma}=\frac{\dot{\rho}}{\ell_B\omega_c}$, and $\ddot{\gamma}=\frac{\ddot{\rho}}{\ell_B\omega_c^2}$ and obtain 
	\begin{equation}
	\label{eq:v1-lagr-rel-nu-1}
	\begin{split}
	\mathcal{L}_{rel}&=\frac{1}{4}\Big[\dot{\gamma}^2+\gamma^2\Big(\frac{\dot{\theta}_{rel}^2}{\omega_c^2}+\frac{\dot{\theta}_{rel}}{\omega_c}\Big)\Big] \\
	&-\frac{a}{q^2\gamma}-\frac{b}{q^2\ell_B\gamma^2}\\
	&-\frac{c\ell_B^2\gamma}{q^2}-\frac{d\ell_B^3\gamma^2}{q^2}
	\end{split}
	\end{equation}
	Since $a=q^2$, we can simplify the above to
	\begin{equation}
	\label{eq:v1-lagr-rel-nu-2}
	\begin{split}
	\mathcal{L}_{rel}&=\frac{1}{4}\Big[\dot{\gamma}^2+\gamma^2\Big(\frac{\dot{\theta}_{rel}^2}{\omega_c^2}+\frac{\dot{\theta}_{rel}}{\omega_c}\Big)\Big] \\
	&-\frac{1}{\gamma}-\frac{B}{\gamma^2}-\Gamma\gamma-\Delta\gamma^2
	\end{split}
	\end{equation}
	with dimensionless quantities $B=\frac{b}{q^2\ell_B}$, $\Gamma=\frac{c\ell_B^2}{q^2}$, and $\Delta=\frac{d\ell_B^3}{q^2}$. With it, we derive
	the following Hamiltonian
	\begin{equation}
	\label{eq:v1-ham-rel-nu}
	\begin{split}
	\mathcal{H}_{rel}&=\frac{1}{4}\Big[\dot{\gamma}^2+\gamma^2(\frac{\dot{\theta}_{rel}^2}{\omega_c^2})\Big] \\
	&+\frac{1}{\gamma}+\frac{B}{\gamma^2}+\Gamma\gamma+\Delta\gamma^2
	\end{split}
	\end{equation}
	the following Newton equations
	\begin{equation}
	\label{eq:v1-newton-rel-nu-1}
	\begin{split}
	&\gamma^4\Big(\frac{\dot{\theta}_{rel}}{\omega_c}+\frac{\dot{\theta}_{rel}^2}{\omega_c^2}-8\Delta-\frac{1}{4} \Big)\\
	&+\gamma^3(-\ddot{\gamma}-6\Gamma)\\
	&+\gamma^2(4\mathcal{H}_{rel}-\dot{\gamma}^2+2p_{\theta}^{rel})\\
	&+\gamma(-2)-4{p_{\theta}^{rel}}^2=0
	\end{split}
	\end{equation}
	and
	\begin{equation}
	\label{eq:v1-newton-rel-nu-2}
	\frac{2\Lambda\dot{\gamma}\dot{\theta}_{rel}}{\omega_c}+\frac{\Lambda^2\ddot{\theta}_{rel}}{\omega^2}+\Lambda\dot{\gamma}=0
	\end{equation}
	where
	\begin{equation}
	\label{eq:bigstar-v1-newton-rel-nu-2}
	\begin{split}
	\Lambda&=\gamma^4\Big(-\Delta-\frac{1}{16}\Big)+\gamma^3(-\Gamma)\\
	&+\gamma^2\Big(\mathcal{H}_{rel}-\frac{\dot{\gamma}^2}{4}+\frac{p_{\theta}^{rel}}{2} \Big)\\
	&-{p_{\theta}^{rel}}^2-B
	\end{split}
	\end{equation}
	and 
	\begin{equation}
	\label{eq:p_theta-v1}
	p_\theta^{rel}=\frac{\gamma^2}{2}\Big(\frac{\dot{\theta}_{rel}}{\omega_c} + \frac{1}{2}\Big)
	\end{equation}
	As it is for $p_\theta^{cm}$, $p_\theta^{rel}$ is a constant of the motion since $\theta$ does not appear in $\mathcal{L}_{rel}$.\footnote{Also, $\{p_\theta^{rel},\mathcal{H}_{rel}\} = 0$.}
	Also, we can see that eqn. (\ref{eq:p_theta-v1}) does not depend on any of the potential constants and thus is the same for all three potentials. 
	\newline
	
	One shall notice that eqn. (\ref{eq:v1-ham-rel-nu}) can be rewritten as a function of $p_\theta^{rel}$ as follows
	\begin{equation}
	\label{eq:v1-ham-rel-p_theta_nu}
	\mathcal{H}_{rel}=\frac{1}{4}\dot{\gamma}^2+\Big(\frac{p_\theta^{rel}}{\gamma} - \frac{\gamma}{4} \Big)^2+V
	\end{equation}
	This form will be useful to calculate the effective potential as it will be explained further in this section. Now, with eqns. (\ref{eq:p_theta-v1}) and (\ref{eq:v1-ham-rel-p_theta_nu}), we can write the following integral of motion
	\begin{equation}
	\label{eq:int-v1}
	\Delta\theta_{rel}=4 \mathlarger{\mathlarger{\int_{\gamma_{min}}^{\gamma}}} \frac{(\frac{p_\theta^{rel}}{\gamma'} - \frac{\gamma'}{4})}{\sqrt{G_1}}d\gamma'
	\end{equation}
	with $G_1$ being the fourth-degree polynomial
	\begin{equation}
	\label{G1}
	\begin{split}
	G_1&=\gamma^4(-1-16\Delta)+\gamma^3(-16\Gamma)\\
	&+\gamma^2(16\mathcal{H}_{rel}+8p_{\theta}^{rel})+\gamma(-16)\\
	&+(-16B-16{p_{\theta}^{rel}}^2)
	\end{split}
	\end{equation}
	We can notice that this integral is in fact an elliptic integral and that no closed form expression of it exists. \newline 
	
	Eqn. (\ref{G1}) turns out to be a fundamental tool to the study of the turning points. If the equation $G_1 = 0$ permits two real and non-negative solutions, then these are the boundaries of our motion that we will call $\gamma_{min}$ and $\gamma_{max}$. Since $G_1$ is a fourth degree polynomial, one can always compute its roots explicitely using Ferrari's method. Also we could have found $G_1=0$ by setting $\dot{\gamma} = 0$ in eqn. (\ref{eq:v1-ham-rel-p_theta_nu}), since this is equivalent to doing $\mathcal{H}_{rel}=V_{eff}$, with $V_{eff}$ being the effective potential. Indeed, the total kinetic energy $T_{tot}$ can be written as a function of $\gamma$ and $\dot{\gamma}$ and is in fact
	\begin{equation}
	\label{eq:T_tot}
	\begin{split}
	T_{tot}(\dot{\gamma}, \gamma)&=T(\dot{\gamma})+T(\gamma)\\
	&=\frac{\dot{\gamma}^2}{4}+\Big(\frac{p_{\theta}^{rel}}{\gamma} - \frac{\gamma}{4} \Big)^2
	\end{split}
	\end{equation}
	Hence, by substituting the conserved angular momentum (\ref{eq:p_theta-v1}) for the angular velocity, we can get rid of time dependence in the effective potential so that it only depends on the radial distance. This implies that eqn. (\ref{eq:v1-ham-rel-p_theta_nu}) is in fact the sum of a kinetic energy term and the effective potential
	\begin{equation}
	\label{eq:v1-ham-t_tot}
	\mathcal{H}_{rel}=T(\dot{\gamma})+V_{eff}=\frac{\dot{\gamma}^2}{4}+V_{eff}
	\end{equation}
	where
	\begin{equation}
	\label{eq:v_eff}
	V_{eff}=T(\gamma)+V(\gamma)
	\end{equation}
	We then obtain the following general relation
	\begin{equation}
	\label{eq:g1-v_eff}
	\mathcal{H}_{rel}-V_{eff}=G=0
	\end{equation}
	valid for all three potentials.
	\newline
	
	Deriving $V_{eff}$ is important to study the behaviour of our particles. In fact, it will help us to determine a particular set of values of the potential's parameters for which we obtain a bounded motion. For these specific constants, all three effective potentials tend to infinity as $\gamma$ approaches $0$ and $\infty$\footnote{One shall remark that this set of constants is a subset of all possible values producing positive turning points for the $G$ polynomials and it was chosen because of its physically intuitive meaning, that is, the particle is always bounded regardless of its energy and thus its position.}. However, since $V_{eff} \leq \mathcal{H}_{rel} \in \mathbb{R}$, there must exist two positive numbers, $\gamma_{min}$ and $\gamma_{max}$, such that
	\begin{equation}
	\label{eq:gamma-min-max}
	\gamma_{min} \leq \gamma(t) \leq \gamma_{max }\text{ for all } t 
	\end{equation}
	
	$V_1$ is composed of the Coulomb potential and the harmonic oscillator. The parameters $B, \Gamma,$ and $\Delta$ can be divided into three categories \footnote{For $B,\Gamma,\Delta>0$.}
	\begin{itemize}
		\item $B$ as a repulsive term which dominates the potential as $\gamma \rightarrow 0$
		\item $\Delta$ as an attractive term which dominates the potential as $\gamma \rightarrow +\infty$
		\item $\Gamma$ as a constant term which acts as a vertical shift for the force
	\end{itemize}
	We shall remark that in the case where $B,\Delta<0$, their role changes. If we analyze the force acting on the particle
	\begin{equation}
	\label{eq:force-v1}
	\begin{split}
	F&\propto-\frac{\partial V_{eff}}{\partial\gamma}\simeq\Big(\frac{p_{\theta}^2+B}{\gamma^3}\Big)\\
	&+\frac{1}{\gamma^2}-\Gamma-\gamma\Big(\frac{1}{16}+\Delta \Big)
	\end{split}
	\end{equation}
	we can explore the behaviour of $F$ as $\gamma \rightarrow 0^+$ and $\gamma \rightarrow +\infty$ to obtain conditions for bounded motion.
	\begin{itemize}
		\item[1.] $\gamma \rightarrow 0^+$: Defining the lower bound $\gamma_{min}$
	\end{itemize}
	It is easy to notice that near $0$, $\frac{p_{\theta}^2+B}{\gamma^{3}}$ is the dominant term of $F$ as we can see with the following limit
	\begin{equation}
	\label{eq:limit-ratio}
	\lim_{\gamma \rightarrow 0^+} \frac{F}{(\frac{p_\theta^{2}+B}{\gamma^3})} = 1
	\end{equation}
	$\implies$
	
	\begin{equation}
	\label{eq:limit asymptotic}
	F \sim \frac{p_\theta^{2}+B}{\gamma^{3}}
	\end{equation}
	In order to have a repulsive force near $0$, we must have
	\begin{equation}
	\label{eq:constant-1-v1}
	B>-p_{\theta}^2
	\end{equation}
	This is the condition for a lower bound of the orbit.

	\begin{itemize}
		\item[2.] $\gamma \rightarrow +\infty$: Defining the upper bound $\gamma_{max}$
	\end{itemize}
	Also, for large values of $\gamma$, $-\gamma(\frac{1}{16}+\Delta)$ is the dominant term of $F$ as we can see with the following limit
	\begin{equation}
	\label{eq:limit ratio2}
	\lim_{\gamma \rightarrow +\infty} \frac{F}{-\gamma(\frac{1}{16}+\Delta)} = 1
	\end{equation}
	$\implies$
	
	\begin{equation}
	\label{eq:limit asymptotic2}
	F \sim -\gamma(\frac{1}{16}+\Delta)
	\end{equation}
	and since we need an attractive force to confine the motion, we find 
	\begin{equation}
	\label{eq:constant-2-v1}
	\Delta > -\frac{1}{16}
	\end{equation}
	This is the condition for an upper bound of the orbit.
	\newline 
	
	One might wonder if $G_1$ could admit more than two turning points. In fact, it is impossible based on Descartes' rule of signs.
	
	\begin{proof}
		\begin{equation}
		\begin{split}
		G_1&=\gamma^4(-1-16\Delta)+\gamma^3(-16\Gamma)\\
		&+\gamma^2(16\mathcal{H}_{rel}+8p_{\theta}^{rel})+\gamma(-16)\\
		&+(-16B-16{p_{\theta}^{rel}}^2)\\
		&=a_4\gamma^4+a_3\gamma^3+a_2\gamma^2+a_1\gamma+a_0
		\end{split}
		\end{equation}
		Based on eqns. (\ref{eq:constant-1-v1}) and (\ref{eq:constant-2-v1}), $a_0, a_4 < 0$. We can see that $a_1$ is also always negative so the combinations that permit the most change of signs would be $a_2 > 0, a_3 < 0$ or $a_2 < 0, a_3 > 0$. In these two cases, by Descartes' rule of signs, $G_1$ has a maximum of two non-negative roots.\newline
	\end{proof}
	
	Furthermore, it could be interesting to see if there exists a relation of the form $f_1(A, B, \Gamma, \Delta, p_\theta^{rel}, \mathcal{H})=\alpha_1 \in \mathbb{R}$ for which we obtain a periodic motion. According to \cite{Hestenes:1999}, eqn. (\ref{eq:int-v1}) must obey the following criterion
	\begin{equation}
	\label{eq:v1-period-relation}
	2\Delta\theta_{rel}=2 \pi + \varepsilon_\theta
	\end{equation}
	where
	\begin{equation}
	\varepsilon_\theta=2 \pi \alpha_1 \text{, with } \alpha_1 \in \mathbb{Q}
	\end{equation}
	is the deviation from an angular period of $2\pi$. This relation arises when $\gamma$ and $\theta$ have commensurable periods and thus describe together a closed orbit. Hence, in order to simplify the above relation, eqn. (\ref{eq:v1-period-relation}) can be rewritten as
	\begin{equation}
	\begin{split}
	\frac{\Delta \theta_{rel}}{\pi} - 1&= \frac{4}{\pi} \mathlarger{\mathlarger{\int_{\gamma_{min}}^{\gamma_{max}}}} \frac{(\frac{p_\theta^{rel}}{\gamma} - \frac{\gamma}{4})}{\sqrt{G_1}}d\gamma  - 1\\ 
	&= f_1(A, B, \Gamma, \Delta, p_\theta, \mathcal{H}) \\
	&= \alpha_1 \in \mathbb{Q}
	\end{split}
	\end{equation}
	
	With the aid of eqns. (\ref{eq:v1-newton-rel-nu-1}) and (\ref{eq:v1-newton-rel-nu-2}), it is possible to plot different orbits for the relative motion between the two identical particles. Here we present three of them with a brief analysis of the chosen parameters. \footnote{We decided to use $\gamma(0)=\gamma_{min}$ with no initial velocities whatsoever.}
	\begin{figure}[h!]
		\centering
		\resizebox{0.30\textwidth}{!}{
			\includegraphics{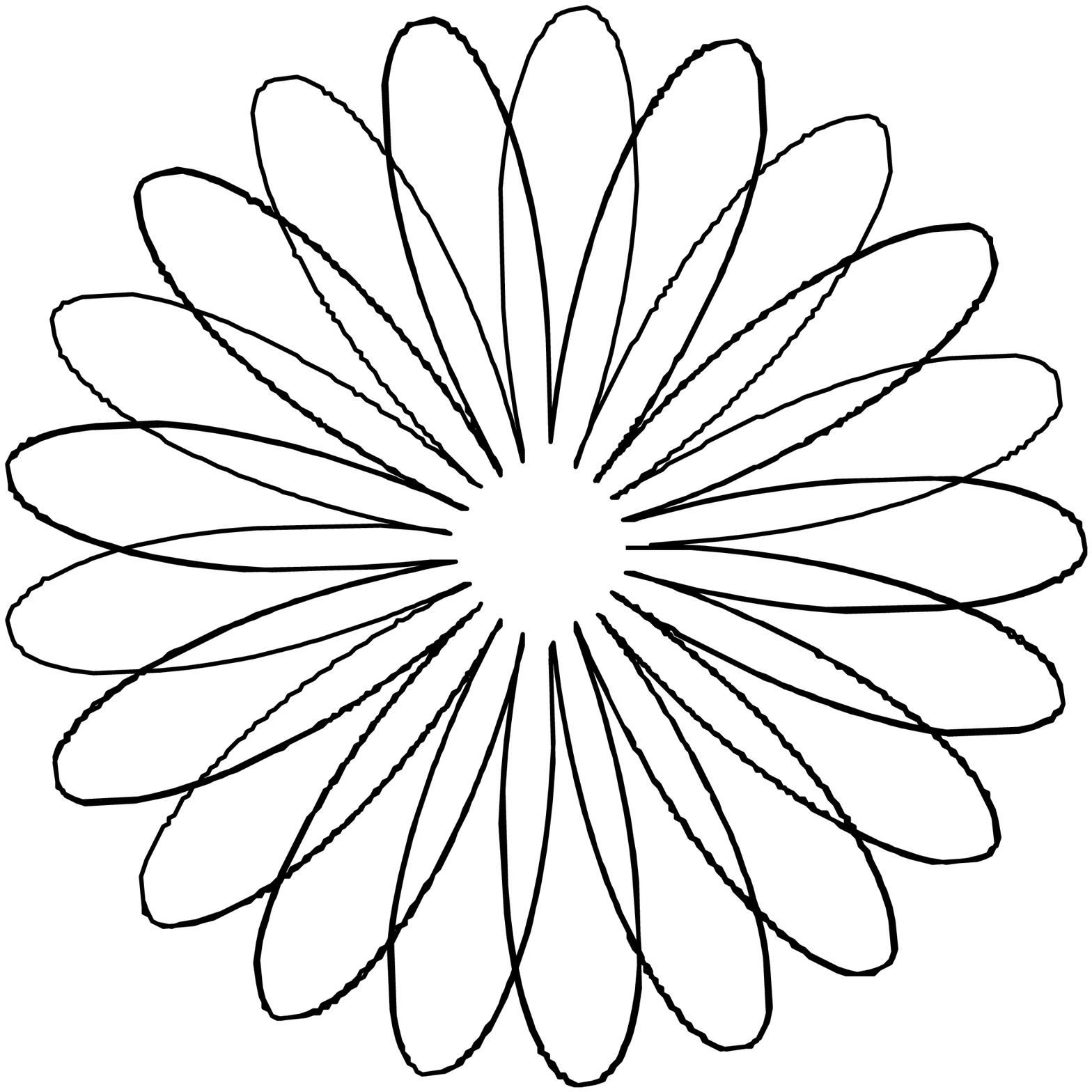}
		}
		\caption{$B=\Gamma=\Delta=1,p_\theta^{rel}=0,\mathcal{H}_{rel}=10$}
		\label{fig:v1_1}
	\end{figure} 
	
	\begin{figure}[h!]
		\centering
		\resizebox{0.30\textwidth}{!}{
			\includegraphics{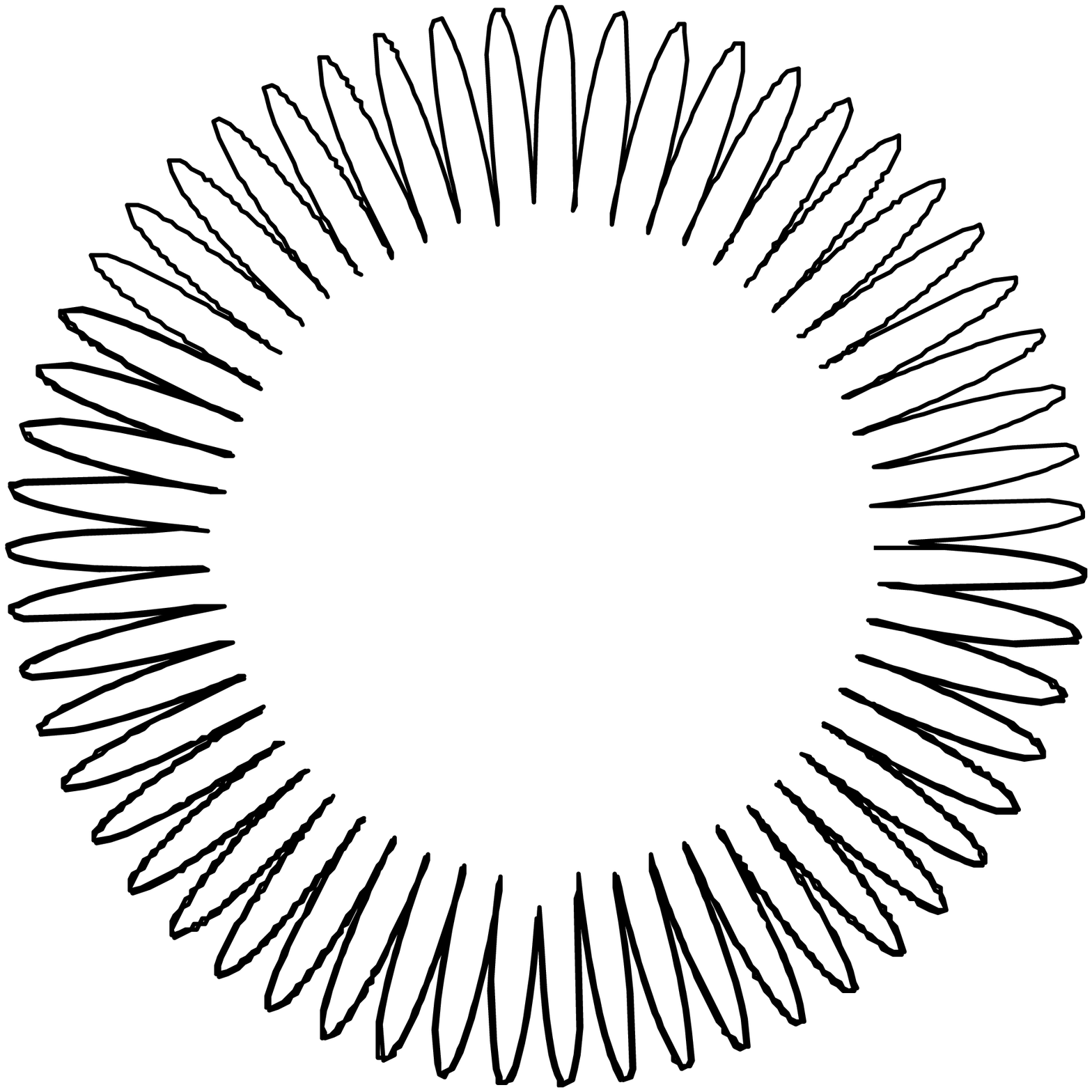}
		}
		\caption{$B=1,\Gamma=\Delta=5,p_\theta^{rel}=0,\mathcal{H}_{rel}=10$}
		\label{fig:v1_2}
	\end{figure} 
	
	\begin{figure}[h!]
		\centering
		\resizebox{0.30\textwidth}{!}{
			\includegraphics{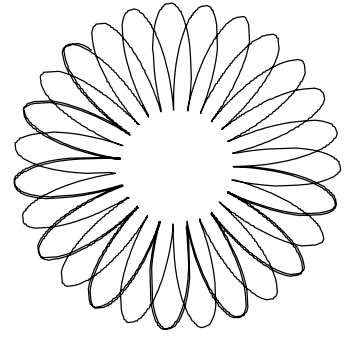}
		}
		\caption{$B=5,\Gamma=\Delta=1,p_\theta^{rel}=0,\mathcal{H}_{rel}=10$}
		\label{fig:v1_3}
	\end{figure} 
	As an example, we use Figure \ref{fig:v1_1} as a base model where all parameters of the potential are the same. In Figure
	\ref{fig:v1_2}, we can see that since we have given domination to the attractive parameters, the particle tends to have smaller oscillation amplitude due to the fact that the force restricts the radial movement. In the case of Figure \ref{fig:v1_3}, we can see that, for the same energy, giving domination to the repulsive parameter increases the value of $\gamma_{min}$
	despite the fact that $\gamma_{max}$ stays roughly the same. Hence, the particle tends to stay away from the center
	while staying in a bounded region since $B$ loses its relevance as $\gamma \rightarrow +\infty$.
	\subsection{Second potential} \label{2nd-potential}
	
	The second potential has the form
	\begin{equation}
	\label{eq:v2-u}
	\boxed{V_2=\frac{a}{\rho^2}+b\rho^2+c\rho^4+d\rho^6}
	\end{equation}
	
	\begin{figure}[h]
		\centering
		\resizebox{0.48\textwidth}{!}{
			\includegraphics{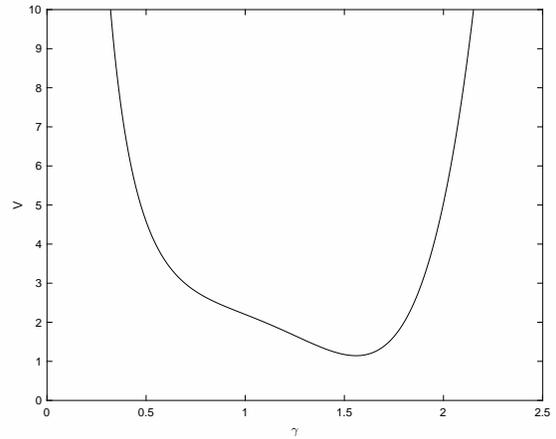}
		}
		\caption{Potential 2}
	\end{figure} 
	
	where $a,b,c,d \in \mathbb{R}$. Again, using dimensionless parameters, we obtain
	\begin{equation}
	\label{eq:v2-lagr-rel-nu-1}
	\begin{split}
	\mathcal{L}_{rel}&=\frac{1}{4}\Big[\dot{\gamma}^2+\gamma^2\Big(\frac{\dot{\theta}_{rel}^2}{\omega_c^2}+\frac{\dot{\theta}_{rel}}{\omega_c}\Big)\Big] \\
	&-\frac{a}{\ell_Bq^2\gamma^2}-\frac{b\ell_B^3\gamma^2}{q^2}\\
	&-\frac{c\ell_B^5\gamma^4}{q^2}-\frac{d\ell_B^7\gamma^6}{q^2}
	\end{split}
	\end{equation}
	
	that can be rewritten as
	\begin{equation}
	\label{eq:v2-lagr-rel-nu-2}
	\begin{split}
	\mathcal{L}_{rel}&=\frac{1}{4}\Big[\dot{\gamma}^2+\gamma^2\Big(\frac{\dot{\theta}_{rel}^2}{\omega_c^2}+\frac{\dot{\theta}_{rel}}{\omega_c}\Big)\Big] \\
	&-\frac{A}{\gamma^2}-B\gamma^2-\Gamma\gamma^4-\Delta\gamma^6
	\end{split}
	\end{equation}
	
	with dimensionless quantities $A=\frac{a}{\ell_Bq^2}$, $B=\frac{b\ell_B^3}{q^2}$, $\Gamma=\frac{c\ell_B^5}{q^2}$, and $\Delta=\frac{d\ell_B^7}{q^2}$. Hence, we can derive the Hamiltonian
	\begin{equation}
	\label{eq:v2-ham-rel-nu}
	\begin{split}
	\mathcal{H}_{rel}&=\frac{1}{4}\Big[\dot{\gamma}^2+\gamma^2\Big(\frac{\dot{\theta}_{rel}^2}{\omega_c^2}\Big)\Big] \\
	&+\frac{A}{\gamma^2}+B\gamma^2+\Gamma\gamma^4+\Delta\gamma^6
	\end{split}
	\end{equation}
	
	and the following integral of motion
	\begin{equation}
	\label{eq:int-v2}
	\Delta\theta_{rel}=4 \mathlarger{\mathlarger{\int_{\gamma_{min}}^{\gamma}}} \frac{(\frac{p_\theta^{rel}}{\gamma'} - \frac{\gamma'}{4})}{\sqrt{G_2}}d\gamma'
	\end{equation}
	
	with $G_2$ being the polynomial
	\begin{equation}
	\label{G2}
	\begin{split}
	G_2&=\gamma^8(-16\Delta)+\gamma^6(-16\Gamma)\\
	&+\gamma^4(-1-16B)\\
	&+\gamma^2(16\mathcal{H}_{rel}+8p_\theta^{rel})\\
	&+(-16A-16{p_\theta^{rel}}^2)
	\end{split}
	\end{equation}
	Unfortunately, using Risch algorithm - based on Liouville's theorem in differential algebra -, we find that this integral has no closed form expression. \newline 
	
	$G_2$ is particular in the sense that it has only even powers of $\gamma$, thus it can be reduced to a fourth degree polynomial in $X = \gamma^2$. By setting eqn. (\ref{G2}) to $0$, the two real and non-negative solutions of the equation are $\gamma_{min}$ and $\gamma_{max}$. As in $G_1$, these can be found in terms of radicals using Ferrari's method.\newline
	
	At the opposite of $V_1$, $V_2$ is peculiar in the sense that it does not contain the Coulomb potential. It contains, however, a harmonic oscillator. The parameters, this time, can be divided into two categories \footnote{For $A,B,\Gamma,\Delta>0$. Otherwise, as in $V_1$, note that for $A,B,\Gamma,\Delta<0$, their role changes.}
	\begin{itemize}
		\item $A$ as a repulsive term which dominates the potential as $\gamma \rightarrow 0^{+}$
		\item $B,\Delta,\Gamma$ as attractive terms which dominate the potential as $\gamma \rightarrow +\infty$
	\end{itemize}
	
	Then, the force acting upon the particle is
	\begin{equation}
	\label{eq:force-v2}
	\begin{split}
	F&\propto-\frac{\partial V_{eff}}{\partial\gamma} \\
	&\simeq\frac{(p_{\theta}^2+A)}{\gamma^3}-\gamma\Big(\frac{1}{16}+B \Big) \\
	&-\Gamma\gamma^3-\Delta\gamma^5
	\end{split}
	\end{equation}
	
	\begin{itemize}
		\item[1.] $\gamma \rightarrow 0^+$: Defining the lower bound $\gamma_{min}$
	\end{itemize}
	We see that $F$ is dominated by $\frac{p_{\theta}^2+A}{\gamma^3}$ near $0$ by looking at the following limit 
	\begin{equation}
	\label{eq:limit-ratio-1-v2}
	\lim_{\gamma \rightarrow 0^+} \frac{F}{(\frac{p_\theta^{2}+A}{\gamma^{3}})} = 1
	\end{equation}
	$\implies$
	
	\begin{equation}
	\label{eq:limit asymptotic3}
	F \sim \frac{p_\theta^{2}+A}{\gamma^{3}}
	\end{equation}
	Hence, for the force to be repulsive, we must have
	\begin{equation}
	\label{eq:constant-1-v2}
	A>-p_{\theta}^2
	\end{equation} 
	which sets the condition for a lower bound of the orbit.
	\begin{itemize}
		\item[2.] $\gamma \rightarrow +\infty$: Defining the upper bound $\gamma_{max}$
	\end{itemize}
	We see that $F$ is dominated by $-\Delta\gamma^{5}$ for large values of gamma as we can see in the following limit
	\begin{equation}
	\label{eq:limit-ratio-2-v2}
	\lim_{\gamma \rightarrow 0^+}\frac{F}{-\Delta\gamma^{5}} = 1
	\end{equation}
	$\implies$
	
	\begin{equation}
	\label{eq:limit asymptotic4}
	F \sim -\Delta\gamma^{5}
	\end{equation}
	Hence, in order to have an attractive force, we must have
	\begin{equation}
	\label{eq:constant-2-v2}
	\Delta>0
	\end{equation}
	This sets the condition for an upper bound of the orbit.
	\newline
	
	As for $G_1$, the question of the number of possible turning points arises. Here, we prove that no more than two can exist.
	
	\begin{proof}
		\begin{equation}
		\label{G2-X}
		\begin{split}
		G_2&=X^4(-16\Delta)+X^3(-16\Gamma)\\
		&+X^2(-1-16B)\\
		&+X(16\mathcal{H}_{rel}+8p_\theta^{rel})\\
		&+(-16A-16{p_\theta^{rel}}^2)\\
		&=b_4X^4 + b_3X^3 + b_2X^2+ b_1X + b_0
		\end{split}
		\end{equation}
		In order for $G_2(X)$ to have four real non-negative roots, there must exist $X_1, X_2 \in \mathbb{R}^+_*$ such that
		\begin{equation}
		G_2''(X_1) = G_2''(X_2) = 0
		\end{equation}
		By Descartes' rule of signs, there must be four changes of sign in $G_2(X)$. Based on eqns. (\ref{eq:constant-1-v2}) and (\ref{eq:constant-2-v2}), this gives the following unique combination of parameters
		\begin{equation}
		\Gamma < 0, B > -\frac{1}{16}, H > -\frac{p_\theta^{rel}}{2}, A > -p_\theta^{{rel}^2}
		\end{equation}
		
		From
		\begin{equation*}
		G_2''(X) = 0
		\end{equation*}
		we find
		\begin{equation}
		\begin{split}
		X_1&=\frac{96\Gamma + \sqrt{Det(G_2''(X))}}{-384\Delta}\\ \\
		X_2&=\frac{96\Gamma - \sqrt{Det(G_2''(X))}}{-384\Delta}
		\end{split}
		\end{equation}
		where
		\begin{equation}
		\begin{split}
		Det(G_2''(X))&=9216\Gamma^2\\
		&+1536\Delta(1-16B)
		\end{split}
		\end{equation}
		Clearly $X_2 > 0$ and, by hypothesis, we obtain
		\begin{equation}
		\begin{split}
		X_1&=\frac{96\Gamma + \sqrt{Det(G_2''(X))}}{-384\Delta} > 0\\
		&\Leftrightarrow 9216\Gamma^2 < 9216\Gamma^2 + 1536\Delta(-1-16B)\\
		&\Leftrightarrow 0 < 1536\Delta(-1-16B) < 0
		\end{split}
		\end{equation}
		from which we derive a contradiction. 
		\newline
		Remark that the case of three turning points is impossible because three changes of sign in eqn. (\ref{G2-X}) cannot occur.\\
	\end{proof}
	
	Again, by using the same process as in $V_1$, one can derive the following relation for periodic motion 
	\begin{equation}
	\begin{split}
	\frac{\Delta \theta_{rel}}{\pi} - 1&= \frac{4}{\pi} \mathlarger{\mathlarger{\int_{\gamma_{min}}^{\gamma_{max}}}} \frac{(\frac{p_\theta^{rel}}{\gamma} - \frac{\gamma}{4})}{\sqrt{G_2}}d\gamma  - 1\\ 
	&= f_2(A, B, \Gamma, \Delta, p_\theta^{rel}, \mathcal{H}) \\
	&= \alpha_2 \in \mathbb{Q}
	\end{split}
	\end{equation}

	To conclude, we present three possible orbits obtained from derived Newton equations
	\begin{figure}[h!]
		\centering
		\resizebox{0.30\textwidth}{!}{
			\includegraphics{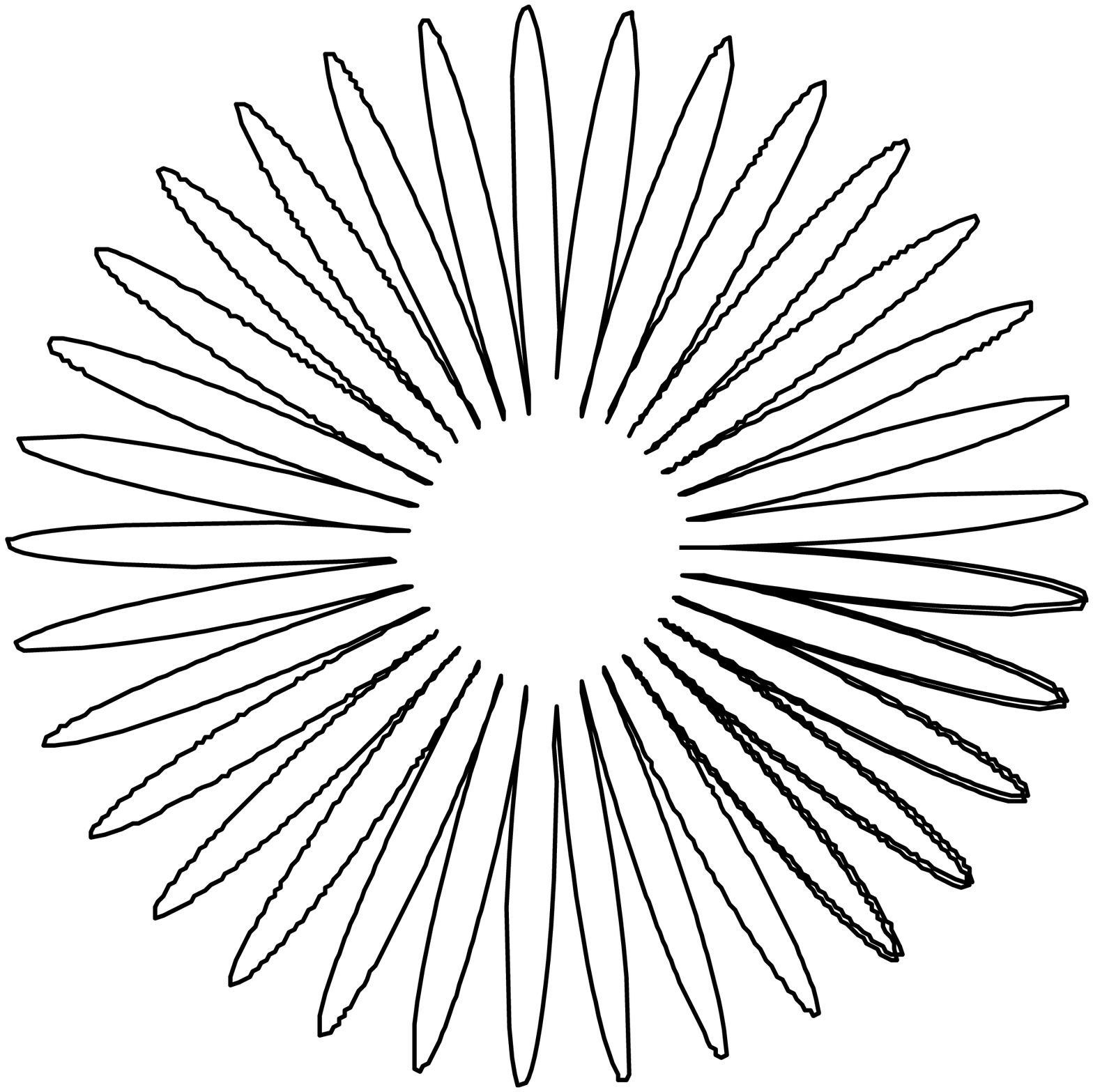}
		}
		\caption{$A=B=\Gamma=\Delta=1,p_\theta^{rel}=0,\mathcal{H}_{rel}=10$}
		\label{fig:v2_1}
	\end{figure} 
	
	\begin{figure}[h!]
		\centering
		\resizebox{0.30\textwidth}{!}{
			\includegraphics{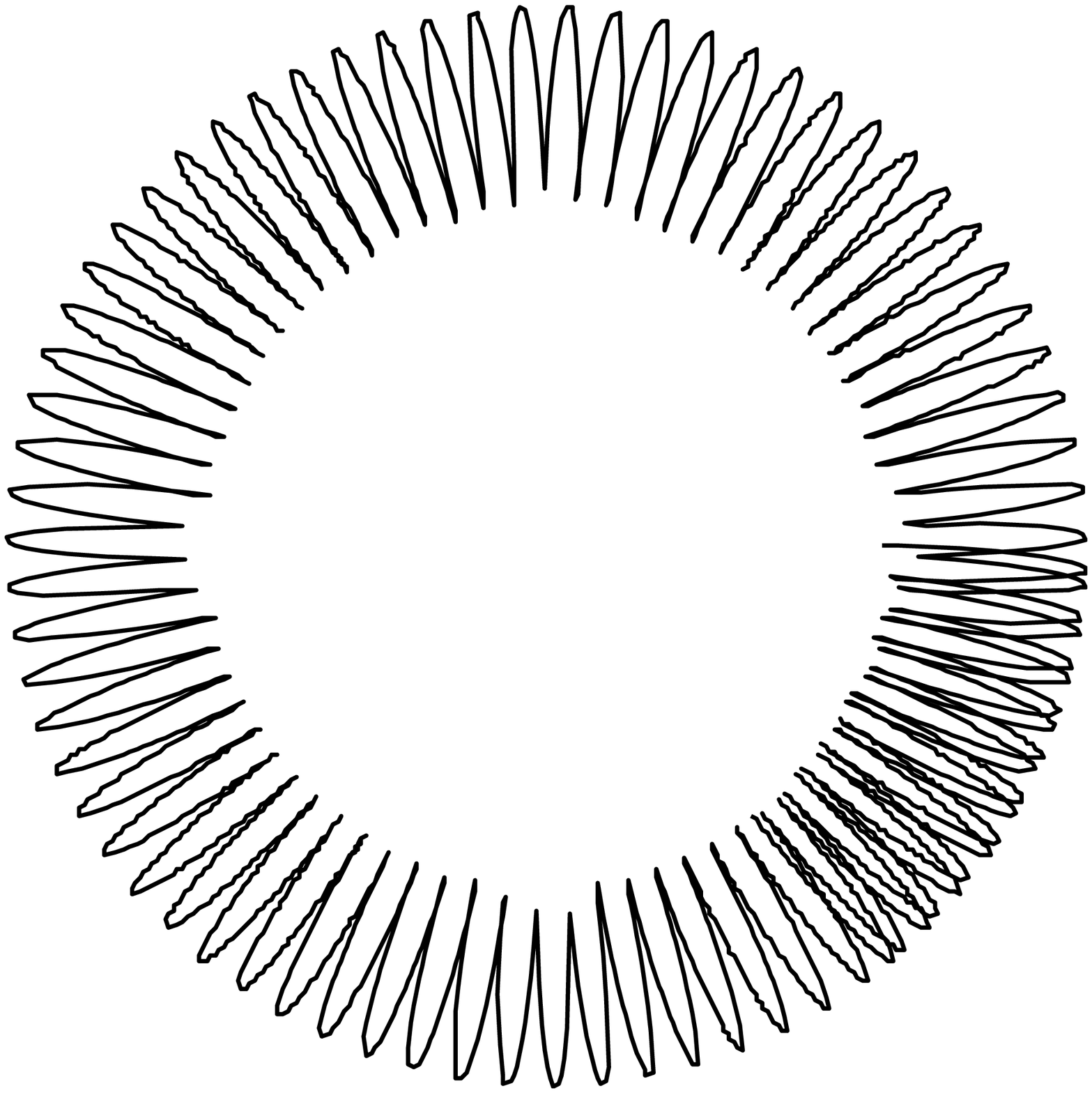}
		}
		\caption{$A=5,B=\Gamma=\Delta=1,p_\theta^{rel}=0,\mathcal{H}_{rel}=10$}
		\label{fig:v2_2}
	\end{figure} 
	
	\begin{figure}[h!]
		\centering
		\resizebox{0.30\textwidth}{!}{
			\includegraphics{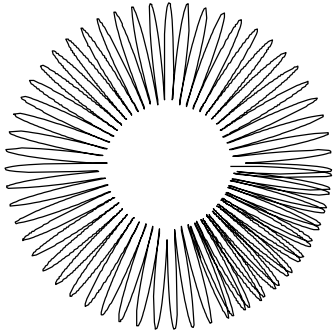}
		}
		\caption{$A=1,B=\Gamma=\Delta=5,p_\theta^{rel}=0,\mathcal{H}_{rel}=10$}
		\label{fig:v2_3}
	\end{figure} 
	
	Since the graph of $V_2$ is similar to the one of $V_1$, we guess that the orbits should have the same shape but thinner and smaller loops because of a steeper curve of the potential. With Figure \ref{fig:v2_1} as a base model, we see that the particle in Figure \ref{fig:v2_2} oscillates farther from the center with a smaller amplitude and in Figure \ref{fig:v2_3} it oscillates in narrower loops.
	\subsection{Third potential} \label{3rd-potential}
	
	The third and last potential has the form
	\begin{equation}
	\label{eq:v3-u}
	\boxed{V_3=\frac{a}{\rho^4}+\frac{b}{\rho^3}+\frac{c}{\rho^2}+\frac{d}{\rho}-e\rho^2}
	\end{equation}
	
	\begin{figure}[h]
		\centering
		\resizebox{0.48\textwidth}{!}{
			\includegraphics{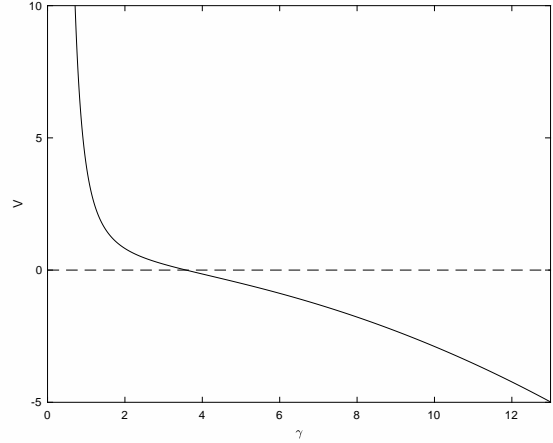}
		}
		\caption{Potential 3}
	\end{figure} 
	where $a,b,c,d,e \in \mathbb{R}$. From it we find
	\begin{equation}
	\label{eq:v3-lagr-rel-nu-1}
	\begin{split}
	\mathcal{L}_{rel}&=\frac{1}{4}\Big[\dot{\gamma}^2+\gamma^2\Big(\frac{\dot{\theta}_{rel}^2}{\omega_c^2}+\frac{\dot{\theta}_{rel}}{\omega_c}\Big)\Big] \\
	&-\frac{a}{\ell_B^3q^2\gamma^4}-\frac{b}{\ell_B^2q^2\gamma^3}\\
	&-\frac{c}{\ell_Bq^2\gamma^2}-\frac{d}{q^2\gamma}+\frac{e\ell_B^3\gamma^2}{q^2}
	\end{split}
	\end{equation}
	that can be rewritten as
	\begin{equation}
	\label{eq:v3-lagr-rel-nu-2}
	\begin{split}
	\mathcal{L}_{rel}&=\frac{1}{4}\Big[\dot{\gamma}^2+\gamma^2\Big(\frac{\dot{\theta}_{rel}^2}{\omega_c^2}+\frac{\dot{\theta}_{rel}}{\omega_c}\Big)\Big] \\
	&-\frac{A}{\gamma^4}-\frac{B}{\gamma^3}\\
	&-\frac{\Gamma}{\gamma^2}-\frac{\Delta}{\gamma}+E\gamma^2
	\end{split}
	\end{equation}
	with dimensionless quantities $A=\frac{a}{\ell_B^3q^2}$, $B=\frac{b}{\ell_B^2q^2}$, $\Gamma=\frac{c}{\ell_Bq^2}$, $\Delta=\frac{d}{q^2}$, and $E=\frac{e\ell_B^3}{q^2}$. Again we derive the Hamiltonian
	\begin{equation}
	\label{eq:v3-ham-rel-nu}
	\begin{split}
	\mathcal{H}_{rel}&=\frac{1}{4}\Big[\dot{\gamma}^2+\gamma^2\Big(\frac{\dot{\theta}_{rel}^2}{\omega_c^2}\Big)\Big] \\
	&+\frac{A}{\gamma^4}+\frac{B}{\gamma^3}\\
	&+\frac{\Gamma}{\gamma^2}+\frac{\Delta}{\gamma}-E\gamma^2
	\end{split}
	\end{equation}
	and the following integral of motion
	\begin{equation}
	\label{eq:int-v3}
	\Delta\theta_{rel}=4 \mathlarger{\mathlarger{\int_{\gamma_{min}}^{\gamma}}} \gamma' \frac{(\frac{p_\theta^{rel}}{\gamma'} - \frac{\gamma'}{4})}{\sqrt{G_3}}d\gamma'
	\end{equation}
	with $G_3$ being the polynomial
	\begin{equation}
	\label{G3}
	\begin{split}
	G_3&=\gamma^6(16E-1)+\gamma^4(8p_\theta^{rel}+16\mathcal{H}_{rel})\\
	&+\gamma^3(-16\Delta)+\gamma^2(-16\Gamma-16{p_\theta^{rel}}^2)\\
	&+\gamma(-16B)-16A
	\end{split}
	\end{equation}
	Here, the use of the Risch algorithm furnishes the proof that no closed form expression of this integral can be found. \newline
	
	$V_3$ is by far the most peculiar potential of this study. Indeed, all terms lead to a repulsive force for $A,B,\Gamma,\Delta,E>0$. By analyzing the behaviour of $V_3$, one might wrongly conclude that, because there is no local minimum, no orbits should exist. Let's push the analysis further. Recall that 
	\begin{equation}
	\label{eq:v3-v_eff}
	\begin{split}
	V_{eff}&=\Big(\frac{p_{\theta}^{rel}}{\gamma} - \frac{\gamma}{4} \Big)^2+\frac{A}{\gamma^4}\\
	&+\frac{B}{\gamma^3}+\frac{\Gamma}{\gamma^2}+\frac{\Delta}{\gamma}-E\gamma^2
	\end{split}
	\end{equation}
	From it, we can find the force
	\begin{equation}
	\label{eq:force-v3}
	\begin{split}
	F&\propto-\frac{\partial V_{eff}}{\partial\gamma}\simeq-\gamma\Big(\frac{1}{16} - E \Big)\\
	&+\frac{A}{\gamma^5}+\frac{B}{\gamma^4}+\Big(\frac{\Gamma+p_{\theta}^2}{\gamma^3} \Big) + \frac{\Delta}{\gamma^2}
	\end{split}
	\end{equation}
	\begin{itemize}
		\item[1.] $\gamma \rightarrow 0^+$: Defining the lower bound $\gamma_{min}$
	\end{itemize}
	We see that $F$ behaves like
	$\frac{A}{\gamma^5}$ for small values of $\gamma$ by evaluating the following limit
	\begin{equation}
	\label{eq:limit-ratio-1-v3}
	\lim_{\gamma \rightarrow 0^+} \frac{F}{(\frac{A}{\gamma^{5}})} = 1
	\end{equation}
	$\implies$
	
	\begin{equation}
	\label{eq:limit asymptotic5}
	F \sim \frac{A}{\gamma^{5}}
	\end{equation}
	Hence for the force to be repulsive, we must have 
	\begin{equation}
	\label{eq:constant-1-v3}
	A>0
	\end{equation} 
	which sets the condition for a lower bound of the orbit.
	\begin{itemize}
		\item[2.] $\gamma \rightarrow +\infty$: Defining the upper bound $\gamma_{max}$
	\end{itemize}
	We see that $F$ behaves like $-\gamma(\frac{1}{16}-E)$ for large values of $\gamma$ by evaluating the following limit
	\begin{equation}
	\label{eq:limit-ratio-1-v3}
	\lim_{\gamma \rightarrow 0^+} \frac{F}{-\gamma(\frac{1}{16}-E)} = 1
	\end{equation}
	$\implies$
	
	\begin{equation}
	\label{eq:limit asymptotic6}
	F \sim -\gamma(\frac{1}{16}-E)
	\end{equation}
	Hence, for the force to be attractive, we must have 
	\begin{equation}
	\label{eq:constant-2-v3}
	E<\frac{1}{16}
	\end{equation} 
	which sets the condition for an upper bound of the orbit.
	On the opposite, for $E>\frac{1}{16}$, we see that the force is repulsive for any $\gamma > 0$. Thus, no orbits can exist since the particle goes away from the center more and more rapidly. At last, there is an interesting case where $E$ is exactly $\frac{1}{16}$. Starting at a certain $\gamma$, the particle is subject to no external force and continues on a constant motion diverging slowly to infinity. Different plots of these cases will be shown at the end of this section.
	\newline
	
	We have seen before that for $V_1$ and $V_2$, the $G$ polynomials can both be expressed as a fourth degree polynomial for which analytic solutions can be computed explicitly. However, this is not the case for arbitrary polynomial equations. In algebra, the Abel-Ruffini theorem, also known as Abel's impossibility theorem, states that there is no algebraic solutions to general polynomial equations of degree five or higher with arbitrary coefficients. Fortunately, a method explained in \cite{Kulkarni:2008} let us explicitly compute solutions of particular equations of degree 6. In our case, one example of restrictions for which we can derive algebraic solutions for $G_3=0$ is
	\begin{equation}
	\label{eq:Restriction-Sextic}
	\begin{split}
	A&=\frac{\zeta\Delta^2}{4}\\
	&-\frac{(\Delta{p_\theta^{rel}}\zeta+2\Delta\mathcal{H}_{rel}\zeta+4B)^2}{4\zeta(p_\theta^{rel}+2\mathcal{H}_{rel})^2-64(\Gamma+p_\theta^{rel^{2}})}
	\end{split}
	\end{equation}
	where
	\begin{equation}
	\label{eq:Restriction-Sextic}
	\zeta=\frac{-16}{16E-1}
	\end{equation}
	It naturally follows that one might wonder if the motion for $V_3$ is restricted to two turning points as we have seen in $V_1$ and $V_2$. In fact, this motion is much richer than the previous ones. As we will see later, because $V_3$ contains not only four but five terms, many more combinations of parameters are possible. Indeed, we can go up to four turning points for a particular set of values of $A,B,\Gamma,\Delta,E$ respecting the necessary conditions for bounded motion.\footnote{The proof of the impossibility to obtain five or six turning points is omitted since it follows directly from Descartes' rule of signs.} 
	\newline
	
	Continuing on the same path, we can derive a similar criterion for periodic motion as seen in $V_1$ and $V_2$
	\begin{equation}
	\begin{split}
	&\frac{\Delta \theta_{rel}}{\pi} - 1\\
	&= \frac{4}{\pi} \mathlarger{\mathlarger{\int_{\gamma_{min}}^{\gamma_{max}}}} \gamma \frac{(\frac{p_\theta^{rel}}{\gamma} - \frac{\gamma}{4})}{\sqrt{G_3}}d\gamma  - 1\\ 
	&= f_3(A, B, \Gamma, \Delta,E, p_\theta^{rel}, \mathcal{H}) \\
	&= \alpha_3 \in \mathbb{Q}
	\end{split}
	\end{equation}
	
	Here we present three relevant cases for the conditions $E < \frac{1}{16}$, $E = \frac{1}{16}$, and $E > \frac{1}{16}$. From the previous analysis, the first case should give a bounded motion, while the second and third ones diverging motions.
	\begin{figure}[h!]
		\centering
		\resizebox{0.30\textwidth}{!}{
			\includegraphics{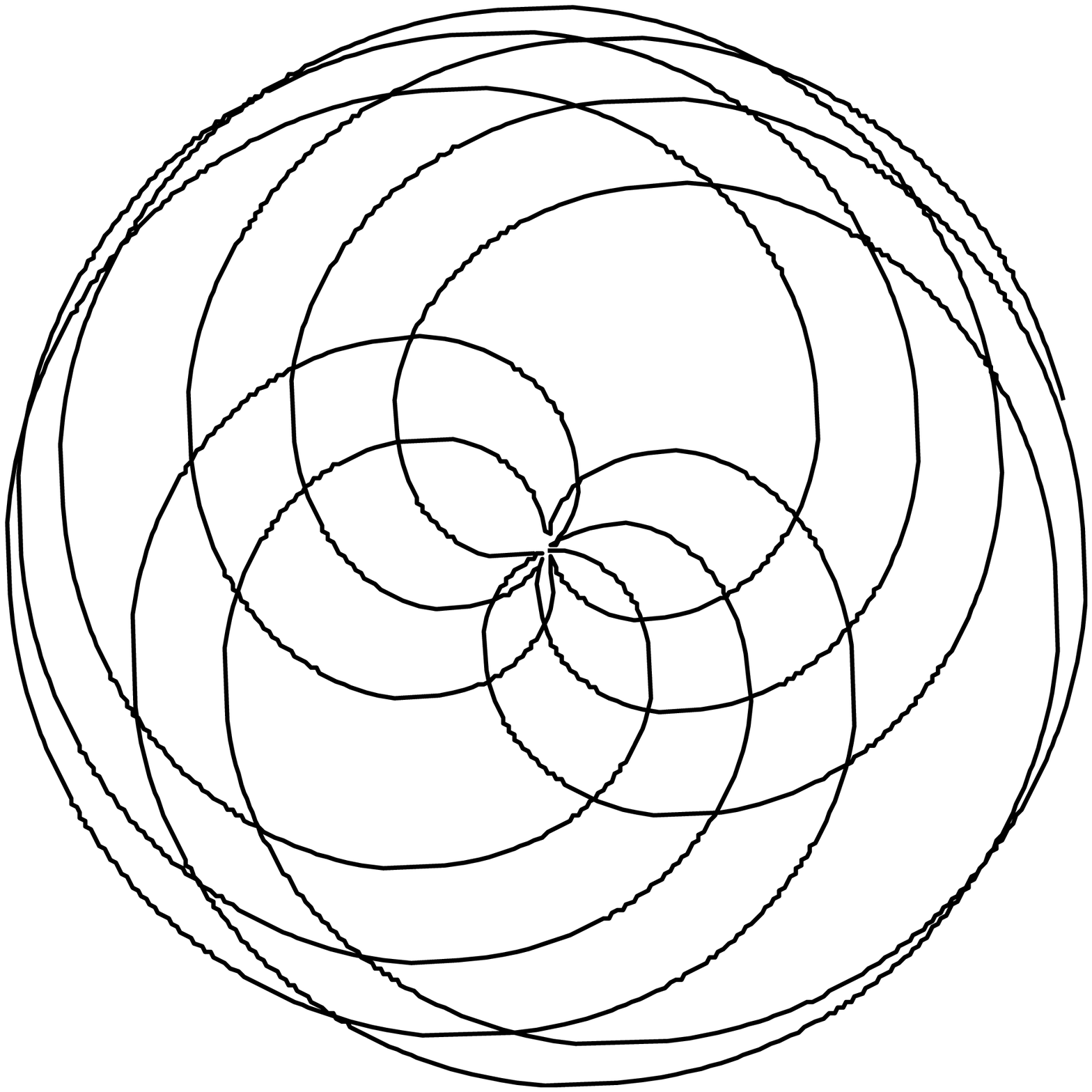}
		}
		\caption{$A,B,\Gamma,\Delta \ll E<\frac{1}{16},p_\theta^{rel}=0,\mathcal{H}_{rel}=10$}
		\label{fig:v3_1}
	\end{figure} 
	\begin{figure}[h!]
		\centering
		\resizebox{0.30\textwidth}{!}{
			\includegraphics{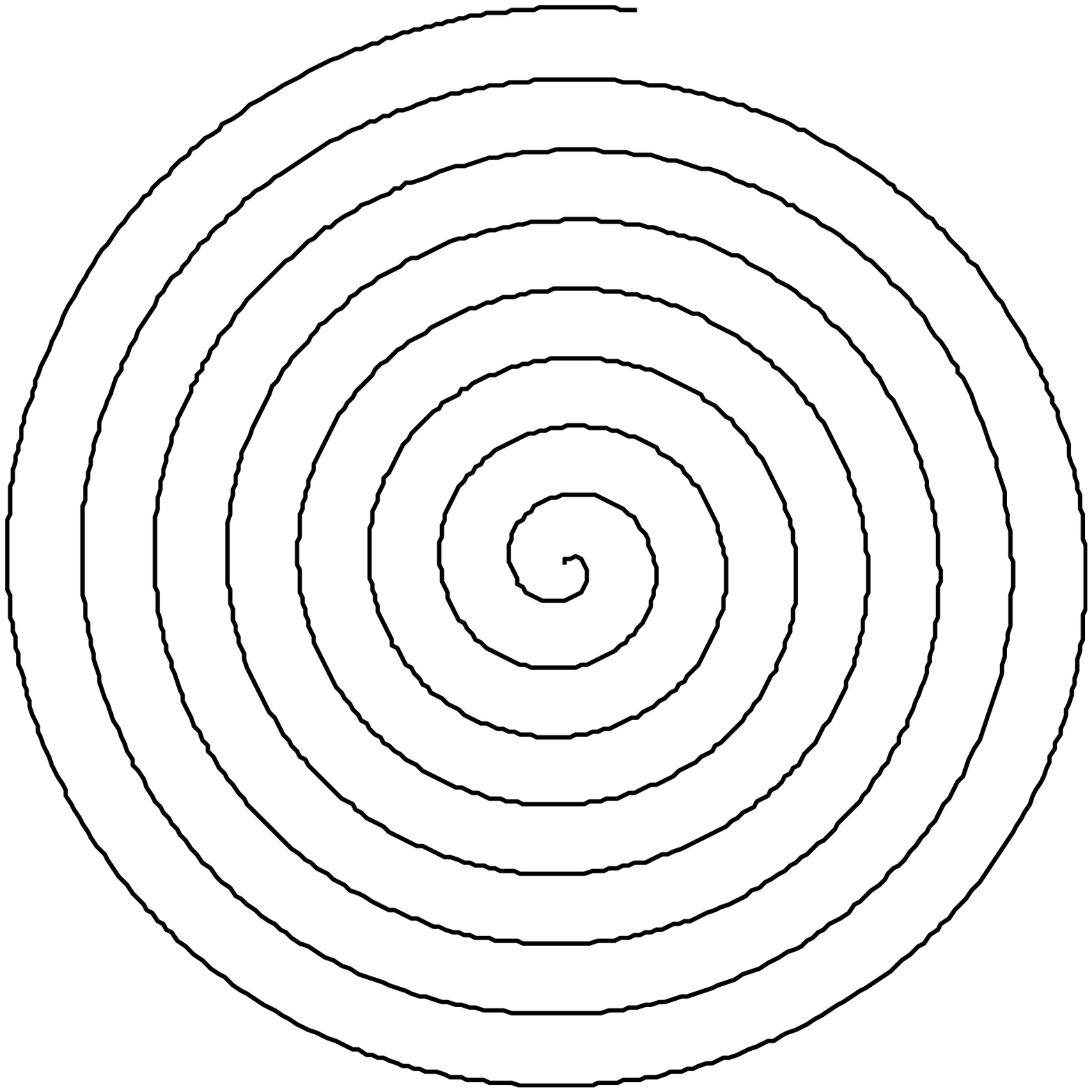}
		}
		\caption{$A=B=\Gamma=\Delta=1,E=\frac{1}{16},\\p_\theta^{rel}=0,\mathcal{H}_{rel}=10$}
		\label{fig:v3_2}
	\end{figure} 
	\begin{figure}[h!]
		\centering
		\resizebox{0.30\textwidth}{!}{
			\includegraphics{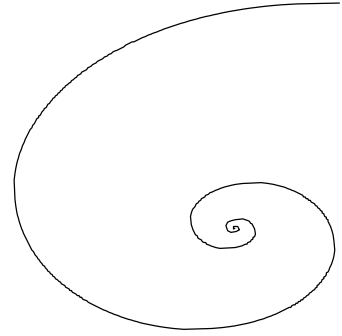}
		}
		\caption{$A=B=\Gamma=\Delta=1,E>\frac{1}{16},\\p_\theta^{rel}=0,H=10$}
		\label{fig:v3_3}
	\end{figure} 
	The following plots show the expected situations for the different cases and agree with our prior analysis. For Figure \ref{fig:v3_1}, the repulsive parameters are quite small compared to $E$ and the condition for bounded motion is respected thus it achieves a conventional orbit. For Figure \ref{fig:v3_2}, however, as $\gamma \rightarrow +\infty$, we can see that $V_{eff} \approx 0$ and thus there is relatively no force being applied on the particle. This is easy to see on the plot by looking at the constant spacing between each circular motion. Finally, as seen in Figure \ref{fig:v3_3}, as soon as the condition for bounded orbits is not respected (\textit{i.e.} $E < \frac{1}{16}$), the potential forces the particle to escape a bounded motion as $\gamma \rightarrow +\infty$ which can be easily seen from the graph by looking at the increasing spacing between each turning of the \textit{spiral}.
	
\begin{figure}[h!]
		\centering
		\resizebox{0.48\textwidth}{!}{
			\includegraphics{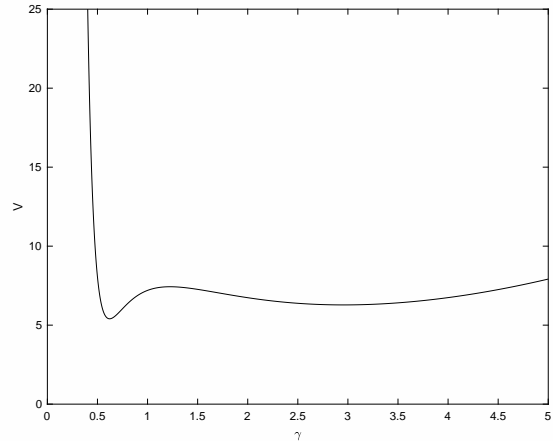}
		}
		\caption{Potential 3 with two wells}
		\label{fig:v3_spec_2}
	\end{figure}

\begin{figure}[h!]
		\centering
		\resizebox{0.3\textwidth}{!}{
			\includegraphics{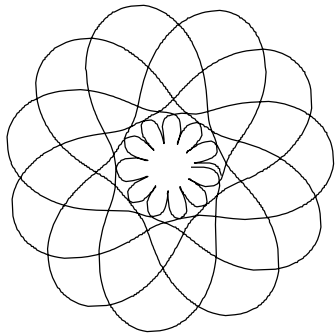}
		}
		\caption{$A=3,B=\Gamma=-6,\Delta=16,E=-0.2$}
		\label{fig:v3_spec_1}
	\end{figure}

	By pushing further the analysis, we can find particular parameters of the potential for which two wells appear as in Figure \ref{fig:v3_spec_2}.
	Depending on the energy of the system, two cases of interest show up. Indeed, the first case is when the energy is set between the local maximum that separates the two wells and the maximum of the minimum of the two wells. We can predict that this yields two distinct motion, thus a case where $G_3=0$ has four non-negative real solutions. The second case is where the energy is above the local maximum. This is an interesting scenario since we can guess the motion would be a mix of two distinct orbits. We can see, in fact, that this is the case in Figure \ref{fig:v3_spec_1}: the particle alternates between two paths. \footnote{Remark that these two paths are not exactly the ones that can be obtained with the same parameters in the first case.}
	\titlespacing{\section}{20pt}{*2}{*0}
	\section{Conclusion}
	Our study was about the classical trajectories in a plane of two identical particles subject to three different non-Coulomb potentials in a constant magnetic field. We have seen that we could separate the problem into two independent problems : the central motion and the relative motion. The center of mass turned out to be moving with a circular motion while the relative motion was more complicated and required further analysis. With the thorough examination of the Hamiltonians and the behaviour of the effective potentials, we have found necessary and sufficient conditions for the existence of a bounded motion and a periodic motion. We have also presented calculations to establish solid basis to the further analysis of a similar case. Indeed, one peculiar and interesting one would be the study of the motion of two quasi-equal particles $(q_1,m_1)$ and $(q_2,m_2)$ obeying the relation $\frac{q_1}{m_1} = \frac{q_2}{m_2}$. Another interesting continuation of the research would be to find if closed-form expressions exist for the integral of motions (\ref{eq:int-v1}), (\ref{eq:int-v2}) and (\ref{eq:int-v3}) by neglecting some terms of the potentials.
	\section{Acknowledgements}
	We would like to thank everybody who helped us to write and complete this article. Especially, we thank P. Winternitz and M.A. Escobar-Ruiz for their special guidance and M. Boukadoum and M. Amir for their meaningful reviews.
	

	

\begin{thebibliography}{99} 
		
		\bibitem[Curilef and Claro, 1995]{CurilefClaro:1995}
		S. Curilef and F. Claro (1995).
		\newblock Dynamics of Two Interacting Particles in a Magnetic Field in Two Dimensions.
		
		\bibitem[Escobar-Ruiz and Turbiner, 2012]{Escobar-RuizTurbiner:2015}
		M.A. Escobar-Ruiz and and A.V. Turbiner (2012).
		\newblock Two charges on plane in a magnetic field: special trajectories.
		
		\bibitem[Escobar-Ruiz and Turbiner, 2013]{Escobar-RuizTurbiner:2013}
		M.A. Escobar-Ruiz and and A.V. Turbiner (2013).
		\newblock Two charges on a plane in a magnetic field: hidden algebra, (particular) integrability, polynomial eigenfunctions.
		
		\bibitem[Hestenes, 1999]{Hestenes:1999}
		David Hestenes (1999).
		\newblock New foundations for classical mechanics
		\newblock pp. 222-223
		
		\bibitem[Kreshchuk, 2015]{Kreshchuk:2015}
		M. Kreshchuk (1995).
		\newblock A quasi-exactly solvable model: two charges in a magnetic field, subject to a non-Coulomb mutual interaction.
		
		\bibitem[Kulkarni, 2008]{Kulkarni:2008}
		R.G. Kulkarni (2008).
		\newblock Solving sextic equations
		\newblock {\em Atlantic Electronic Journal of Mathematics}, 3:56--60.
		
		\bibitem[D. Pinheiro and R. S. Mackay, 2006]{PinheiroMackay:2006}
		D. Pinheiro and R. S. Mackay (2006).
		\newblock Interaction of two charges in a uniform magnetic field: I. Planar problem.
		
	\end{thebibliography}
\end{document}